\documentclass[manuscript]{acmart}

\AtBeginDocument{%
  }

\copyrightyear{2026}
\acmYear{2026}
\acmConference[CHI '26]{Workshop on Engagement in Digital Health Interventions: Open Questions for Research and Design}{April 16, 2026}{Barcelona, Spain}

\begin{document}

\title{Contrasting Perspectives on Engagement Across Three Digital Behavior Change Interventions}

\author{Evangelos Karapanos}

\orcid{0000-0001-5910-4996}
\affiliation{%
  \institution{Cyprus University of Technology}
  \city{Limassol}
  \country{Cyprus}
} 
\email{evangelos.karapanos@cut.ac.cy}

\author{Ruben Gouveia}

\affiliation{%
  \institution{LASIGE, Faculdade de Ciências}
   \institution{Universidade de Lisboa}
  \city{Lisbon}
  \country{Portugal}
}
\email{rhgouveia@ciencias.ulisboa.pt}

\renewcommand{\shortauthors}{Trovato et al.}

\begin{abstract}
We contrast three perspectives on engagement from three projects on the design of Digital Behavior Change Interventions (DBCIs), all conducted as part of the PhD thesis of the second author. We provide a reflection on this work with respect to engagement, discussing the motivation, the assumed effects of engagement, the measures of engagements and key insights of each project, as the well as the strategies employed to increase engagement. 
\end{abstract}



\maketitle

\section{Introduction}
User engagement has become a focal concept in the design of digital behavior change interventions (\textit{e.g.}, \cite{yardley2016understanding, perski2017conceptualising}, not without reason. First, user engagement is a precondition for many of the prominent behavior change techniques to work. For instance, research on self-monitoring has shown that individuals quickly relapse to their old habits once they stop to monitor their behaviors \cite{baumeister1996self}, while recent empirical studies have shown a dose-response association between user engagement with successful behavior change \cite{yardley2016understanding}. Second, early empirical studies, for instance in the context of physical activity tracking, have highlighted high-drop out rates among users of such devices (\textit{e.g.}, \cite{shih2015use}). Third, inquiries into users’ engagement with different intervention components may reveal false assumptions about the use, and, in turn, effects of an intervention component on individuals’ behaviors. 

In this position paper we contrast three perspectives on engagement from three projects on the design of Digital Behavior Change Interventions (DBCIs), all conducted as part of the PhD thesis of the second author. We provide a reflection on this work with respect to engagement, discussing the motivation, the assumed effects of engagement, the measures of engagements and key insights of each project, as the well as the strategies employed to increase engagement. All three projects emphasize a behavioral perspective on engagement - that is the \textit{frequency}, \textit{duration}, \textit{depth} (\textit{i.e.}, variety of content engaged with) of users’ interaction with the technology \cite{yardley2016understanding, perski2017conceptualising}, as well as its \textit{proximal impact} on individuals’ behaviors or behavioral antecedents. 

\section{Three perspectives on Engagement}

\subsection{Habito}
Habito \cite{gouveia2015we} was a mobile application, specially designed and built in order to enable us to study how users engage with physical activity trackers. Habito leveraged two behavior change techniques: \textit{goal setting} and \textit{self-monitoring}.

\textbf{Motivation.} Our work on Habito was motivated by the high drop-out rates of physical activity trackers. Behavior change techniques such as self-monitoring, and goal-setting, require active user engagement for them to be effective.

\textbf{Assumed effects of engagement.}  Latent effects on knowledge, attitudes and motivation. 

\textbf{Measures and insights.}  Three different facets of engagement were measured and analyzed: 1) \textit{Adoption rate}. Users were classified into adopters (\textit{i.e.}, use > 1 week, 38\%) and non-adopters. Stage of change \cite{prochaska1997transtheoretical} was a significant prediction of adoption rate, with people in the intermediary stages - those that have the motivation to change their behaviors but have no developed plans for doing so, having an adoption rate of about 50\%, while individuals in the early (i.e., precontemplation) or late stages (i.e., action and maintenance) having an adoption rate of only 20\%. 2) \textit{Type of usage session.} Usage sessions classified into \textit{glance}, \textit{review} and \textit{engage} based on their duration and interaction nature. We found that 60-70\% of usage sessions to be \textit{glances}  - brief, of median duration 5 seconds, where users only looked at their current activity levels without scrolling to further data. 3) \textit{Re-visitation patterns}. 29\% of usage sessions were separated by less than 5 minutes. Time to next session increased as people progressed towards their daily goal, supporting the view of trackers as "deficit" technologies.

\textbf{Strategies to increase engagement:} \textit{Novelty}. Motivated by prior work on checking habits \cite{oulasvirta2012habits} - "brief, repetitive inspection[s] of dynamic content", we drafted 91 textual messages attempting to contextualize physical activity and explored whether \textit{novelty} (\textit{i.e.}, encountering new messages), thus sustaining the informational reward of usage sessions, would affect re-visitation patterns and sustain engagement. We found that when users encountered a novel message, they were more likely to swipe to see more messages, interacted with Habito for a longer time, and took less time to re-engage with Habito.

\subsection{Glanceable UIs}
In the second project \cite{gouveia2016exploring}, we prototyped and evaluated four Android watchfaces that provide glanceable visualizations of physical activity. For instance, \textit{TickTock} (see fig. portrays, on the periphery of one's watch, periods in which one was physically active over the past hour. Each minute of the past hour is represented by a dash, being white if one was sedentary over the minute, and blue if one was active. These four prototypes were selected out of 

\textbf{Motivation.} If the majority of users' interactions with physical activity feedback are glance sessions, as we found in the Habito project, how can we make physical activity feedback glanceable? An inquiry into the design space of glanceable feedback for physical activity tracking  resulted in 21 unique concepts and 6 design qualities: being \textit{abstract}, \textit{integrating with existing activities}, \textit{supporting comparisons to targets and norms}, being \textit{actionable}, having the capacity to lead to \textit{checking habits }and to act as a \textit{proxy to further engagement}. We then prototyped four of the interfaces and carried out a month-long field study with 12 participants.

\textbf{Assumed effects of engagement.}  Proximal effects on motivation and behavior.

\textbf{Measures and insights.} Engagement and its impact on behavior were explored in the following two ways: 1) \textit{Frequency and nature of engagement}. On average, users glanced at their watch 107 times per day. 41\% of those followed a notification. One of the four UIs was associate with significantly lower engagement and daily steps than the remaining three. 2) \textit{Proximal effects of feedback on behavior}. For each interface, we looked at the proximal impact of each usage session on a users' behaviors, as a function of the feedback users were exposed to. For instance, on \textit{TickTock}, we found that if participants glanced at their watch and saw that they were particularly inactive over the past hour, they were more likely to initiate a new walk in the near future. 

\textbf{Strategies to increase engagement}. In our design exploration we theorized that glanceable physical activity feedback should serve also two objectives with respect to engagement: a) it should be able to sustain user engagement and instigate checking habits, and b) it should lead to "aha" moments and act as a proxy to further engagement. The first objective was explored through two notions: a) \textit{novelty} (\textit{i.e.}, presenting novel content each time a user engages with the UI; as in the CrowdWalk concept, which presents walking activities one can perform around her current location), and b) \textit{scarcity} (\textit{i.e.}, making feedback a scarce resource, as in TickTock, which portrays physical activity feedback only of the past hour. The second objective (\textit{i.e.}, acting as a proxy to further engagement) could be served through providing feedback that raises more questions than provides answers, as in the Meanfull concept, which highlights patterns in user data through textual messages (e.g., "Lazy Tuesdays") while offering the opportunity to further explore the underlying data. 

\subsection{Mikro}
Mikro \cite{gouveia2026breaking} is a watchface that enables users to set \textit{micro-goals} - brief, situated goals on physical activity (\textit{e.g.}, walk 300 meters in the next 10 minutes). 

\textbf{Motivation.} Daily physical activity goals (\textit{e.g.}, 10K steps per day) constitute the norm in physical activity promotion technology. Yet, daily activity goals often fail to adapt to the complexity of daily life and can turn into constant reminders of missed expectations. Building on our prior work \cite{gouveia2018activity} that revealed how individuals establish practices that enable them to leverage small opportunities for walking through opportunistic "micro-plans", we designed \textit{Mikro} as a technology probe to explore why, when, and how people form micro-goals, and how this affects physical activity outcomes. 

\textbf{Assumed effects of engagement.} Proximal effects on behavior as well as an increased salience of physical activity within an individual’s cognitive processes.

\textbf{Measures and insights.} The study aimed at inquiring into how users engaged with Mikro, and what this had on their physical activity. We explored: 1) \textit{Practices and motives for micro-goal setting}. We identified five motives, such as having fun, triggering action, and documenting and learning about activities, among others. We found that participants adjusted the default suggested goal up to 80\% of the times. 2) \textit{Impact of micro-goals of proximal physical activity}. We found that the default duration of goals mattered, with 10-min goals being particularly effective at triggering immediate action, with about 29\% of them being followed by walking within the next minute, while this dropped to about half (9-14\%) for goals of other duration (i.e., 20, 30, or 60 minutes).

\bibliographystyle{ACM-Reference-Format}
\bibliography{sample-base}

\end{document}